\documentstyle[art11]{article}
\def\PRL#1{{ Phys. Rev. Lett.} { #1}}
\def\PRD#1{{ Phys. Rev.} { D#1}}
\def\NPB#1{{ Nucl. Phys.} { B#1}}
\def\PLB#1{{Phys. Lett.} { B#1}} 
\def\vev#1{\langle #1 \rangle}

\font\tenrm=cmr10

\newcommand{\be}{\begin{equation}}
\newcommand{\ee}{\end{equation}}

\begin{document}

\title{\huge     Topological defects and conditions for
     baryogenesis  in the Left-Right symmetric 
     model}

\author{U. A. Yajnik$^a$\thanks{yajnik@niharika.phy.iitb.ernet.in} ,
 Hatem Widyan$^b$ \thanks{hatem@ducos.ernet.in} ,  Shobhit Mahajan$^b$
\thanks{sm@ducos.ernet.in} , Amitabha Mukherjee$^b$
\thanks{am@ducos.ernet.in},\\ and Debajyoti Choudhury$^c$
\thanks {debchou@mri.ernet.in}  \\
{$^a${\tenrm Physics Department, Indian Institute of Technology
Bombay, Mumbai 400\thinspace076, India}} \\
{$^b${\tenrm Department of Physics and Astrophysics, University of
Delhi, Delhi 110 007, India}} \\
$^c$\tenrm Mehta Research Institute of Mathematics and Mathematical 
	Physics,\\ Chhatnag Road, Jhusi, Allahabad 221 506, India}

\date{\today}
\maketitle

\begin{abstract}
{It is shown that the minimal Left-Right symmetric model admits
cosmic string, domain wall and conditionally, monopole solutions. 
The strings arise when the $SU(2)_R$ is broken and can either be 
destabilized at the electroweak scale or remain stable through the 
subsequent breakdown to $U(1)_{EM}$. 
The monopoles and domain wall configurations 
exist in the  $SU(2)_L\otimes U(1)_Y$ symmetric phase and 
disappear after subsequent symmetry breaking.  
Their destabilization provides new sources of non-equilibrium
effects below the electroweak scale. Several defect-mediated 
mechanisms for low energy baryogenesis are 
shown to be realisable in this model.}

\end{abstract}
%\pacs{\tt$\backslash$\string pacs\{12.10.Dm, 98.80.Cq, 98.80.Ft\}}

\section{Introduction}
Spontaneously broken gauge theories typically 
possess topological solutions\cite{coleman}.
The presence of such objects in the early universe 
has been shown to be natural\cite{kibble,vachasrev}. 
Some  of these objects, e.g. cosmic strings, have been 
investigated for their role in structure formation and baryogenesis
\cite{kib,sbdanduay1,rbetal1,garvacbar,rbetal2}. On the 
other hand, monopole and domain wall solutions
\cite{vachasrev,kib,presk,vilenshel} have undesirable 
cosmological consequences, barring a few exceptional circumstances.
The requirement of their absence puts stringent limits on the theory.

 Currently, several unification schemes are being investigated in
detail, specially for their signatures in the 
planned particle accelerators. Some of the unification schemes have
interesting consequences for cosmology. A rich
variety of cosmic string solutions was
demonstrated\cite{asanduay,uayexcon} in the context of
$SO(10)$ unification and has received fresh attention
\cite{acdetal}. 
Furthermore, as the non-viability of several models 
for electroweak baryogenesis is becoming 
apparent\cite{carwag,clijoka},
it is interesting to search for new
mechanisms for low energy baryogenesis in other unified
models\cite{rbetal2,rj1}.

In this paper we investigate the minimal Left-Right symmetric 
model for the presence of topological solutions. 
In section II we discuss the topology of the vacuum sector
of the theory. Two possibilities for the same are distinguished
depending upon the nature of the Higgs potential.
In section III we show that one possibility leads to stable strings and 
no monopoles. The other possibility leads to 
monopoles and metastable cosmic strings.
The monopoles disappear after electroweak symmetry breaking.
The fate of the string depends on several factors, but at
least some are shown to survive to the present epoch. We also show
the existence of zero modes for neutrinos. In section IV we
discuss a domain wall solution, also unstable below the electroweak
scale. All these objects can play an important role in cosmology,
which is discussed in the concluding section.

\section{Topological considerations}
The Left-Right symmetric unification group
$SU(2)_L \otimes SU(2)_R \otimes U(1)_{B-L}$ possesses 
a $U(1)$ whose gauge charge turns out, in a natural and 
compelling way, to be the $B-L$ number of the observed fermions. 
We use the conventions of Mohapatra\cite{mohapbook}, except that
the  $U(1)$ charge assignments are the value of $X=(1/2)(B-L)$.
We begin with the phase in 
which only the first stage of symmetry breaking
$SU(2)_R\otimes U(1)_{B-L}$$\rightarrow$$U(1)_Y$ has occurred.
The field
signalling this breakdown is the $(1,0,1)$ field
$\Delta_R$ which acquires the vacuum expectation
value (vev) with the $(2,1)$ entry of the matrix being the
only non-trivial component, $\langle \Delta_R \rangle_{21}=v_R$.
This choice leads to the $U(1)$ generated by $Y=T^3_R+X$ 
left unbroken. The hypercharge $Y$ is unbroken also by
all other matrices of this form, i.e., 
$\langle \Delta_R \rangle_{21}=v_R\exp\{i \theta\}$ with
$0\le\theta<2\pi$ for a given $v_R$. These therefore constitute
the vacuum manifold. 

If the potential of the $\Delta_R$ field is
such as to allow more general matrices to be possible vevs, 
$Y$ should turn out to be broken, replaced perhaps by some other
generator of the parent group. This can be tolerated
before electroweak symmetry breaking, but conflicts with
phenomenology at lower energies since electric charge
$Q=T^3_L+Y$. The complete potential
involving all the Higgs fields of the theory will therefore be
assumed to tolerate $Q$ preserving vevs only, but the part
involving only the $\Delta_R$, and including temperature
dependent corrections  may either break or preserve $Y$. 
We show in the next section that these two possibilities lead 
to separate interesting results. 

Next, we note that in the parent group, each factor $SU(2)$ and $U(1)$
is multiply connected for the purpose of present considerations. 
The factor $U(1)_{B-L}$ is a compact $U(1)$. This clear because
the $X$ charge of the $Delta_R$ is integer. But more fundamentally,
this is because all known and proposed particles 
carry integer multiples of the basic unit $1/6$ of this charge,
carried by the quarks. This factor is therefore not simply
connected. Secondly, acting on the $\Delta_R$ field,
the $SU(2)$ is effectively an $SO(3)$. This is because
the $\Delta_R$ is a 3-dimensional vector, albeit with complex
components. $SO(3)$ too is not simply connected. The existence
or otherwise of topological objects therefore depends on
new nontrivial closed curves or discrete symmetries appearing
due to the form of the vev.

\section{Cosmic strings and monopoles}
Consider first the case where $Y$ is preserved throughout.
This makes the manifold of inequivalent vacua isomorphic to
$S^1$, a circle. A cosmic string ansatz can be constructed 
by selecting a
map $U^\infty$ from the circle $S^\infty$ at infinity
into some broken $U(1)$ subgroup of the original group,
such that action of this $U(1)$ makes the vev traverse 
the complete manifold of inequivalent vacua. Such a
$U(1)$ is generated by ${\tilde Y}=T^3_R-X$. 
Furthermore, we select the internal 
parameter to be one-half times the spatial cylindrical angle 
$\theta$. Thus,
\be
U^\infty(\theta) = \exp\{i{\theta\over 2}(T^3_R-X)\}
\ee
The $SU(2)$ acts on $\Delta_R$ by similarity transformation, so
$\langle \Delta_R(\infty, \theta) \rangle_{21} = e^{- i\theta}v_R$.
The vev therefore traces the whole $S^1$; however,
\be 
U^\infty(2\pi) = e^{-i\pi}\pmatrix{i &0\cr 0 &-i} 
\neq U^\infty(0) ~.
\ee
Thus we have identified a $Z_2$ which leaves the
vev invariant but not the general matrix $\Delta_R$.
The assumption regarding the form of the
potential has reduced the problem to that of $U(1)_{T^3_R}\otimes
U(1)_{B-L}$ $\rightarrow U(1)_Y$ breaking. This example is easily
analysed to show the existence of topological strings.
The ansatze for the gauge fields are derived from the 
$U^\infty(\theta)$ and the required asymptotic forms for
$r\rightarrow\infty$ are found to be 
\be
W_{R\theta} = {T^3_R\over 4rg_R} \qquad
{\rm and} \qquad B_\theta = {X \over 2rg'}~,
\label{ggeansatz}
\ee
where $g_R$ and $g'$ are the gauge couplings.

Consider next the possibility that the vacuum manifold is
2-dimensional. 
The $Z_2$ identified above makes the stability
group to be $\exp\{i\theta(T^3_R+X)\}$, with $0\le\theta<\pi$.
The restriction of the $U(1)$ parameter to half its natural 
range means a new unshrinkable curve and the kernel
of the natural homomorphism\cite{coleman} of $\Pi^1(H)$ to
$\Pi^1(G)$ becoming non-trivial. Thus monopoles become possible.
Monopole-antimonopole pairs can also form, connected by 
string configurations previously identified.
The string configurations on the other hand lose their
topological stability because the vacuum manifold becomes
isomorphic to $S^2$.

The configurations identified above do not survive the
subsequent phase transition. For analysing low energy vacuum
structure, we must assume only $Q$ preseving vacua.
Whereas monopoles cannot exist, cosmic
strings can be shown to exist by generalising the analysis 
of the first case above. The low energy vevs of the
$(1,0,1)$ field $\Delta_L$ and the $({1 \over 2},{1 \over 2},0)$ 
field $\phi$ are, respectively, $\langle \Delta_L
\rangle_{21}=v_L$ and diag$(\kappa, -\tilde\kappa)$ 
which are not invariant under the action of $U^\infty(2\pi)$.
However, one may think of the above curve $U^\infty(\theta)$
as a projection to the subspace $SU(2)_R\otimes U(1)_{B-L}$
of the more general curve
${\tilde U}^\infty(\theta) = \exp\{i(T^3_R+T^3_L-X)\theta/2\}$.
This leaves $\Delta_R(\infty,\theta)$ to be as above and leaves
the $\phi$ vev invariant, but makes
$\langle \Delta_L(\infty, \theta) \rangle_{21} = e^{i\theta}v_L$.
The argument for stability remains the same as before.

If such cosmic strings form, they should exist as relics
at the present epoch. At the electroweak phase transition,
if the vevs of $\Delta_L$ and $\phi$ in the domains around
an existing vortex are not too different from each other, they will
destabilize the vortex. If the new vevs wind nontrivially in the
internal space while traversing a closed physical path
around the existing vortex, then a stable string forms. 

It may be noted that the stable strings necessarily contain 
$SU(2)_R$ charged condensates. Therefore two possible 
scenarios need to be considered while estimating relic
string density at any epoch. If the vev structure is $Q$
and $Y$ preserving throughout, no new strings arise at the
electroweak phase transition since the latter generically 
does not release latent heat sufficient to excite $SU(2)_R$ 
charged field condensates. On the other hand 
if the potential is such as to allow monopoles above the
electroweak scale, the corresponding strings are at best
metastable. The strings then form only as connecting monopole-
antimonopole pairs and continue to contract and disappear.
However, those monopoles not attached to strings must 
disintegrate at the electroweak phase
transition. This may release large amounts of latent heat,
creating new string-like defects, which can become stable
according to the scenario of the previous paragraph.

The cosmic strings also carry fermion zero modes.
The $\tilde Y$ charge of all the fermions is $\leq 1$.
Hence\cite{nohlveg} the gauge field of the minimal vortex 
will not bind any fermions in a zero-energy mode. 
The Yukawa coupling of the $\nu_R$ and $\nu_L$ for each flavour
contains terms $\psi^T_R C^{-1} \tau_2 \Delta_R \psi_R$ and
$\psi^T_L C^{-1} \tau_2 \Delta_L \psi_L$.
Substituting the ansatz for the Higgs fields, we see that
above the electroweak scale, the $\Delta_R$ term undergoes a $2\pi$ 
phase change around the minimal vortex,  giving rise to solitary\cite{jrew}
zero mode. For the strings surviving the electroweak breaking 
transition, additionally, $\nu_L$ possesses a zero mode for
the same reason.
More zero modes due to both gauge and Higgs coupling are indeed
possible for vortices with higher winding number.  

\section{Domain wall }

At tree level the Lagrangian is symmetric under the exchange
$\Delta_L \leftrightarrow\Delta_R$, reflecting the hypothesis 
of $L-R$ symmetry. If the vacuum values for these
two Higgs fields are assumed to be as in the previous section,
it can be shown \cite{mohapbook} that their 
potential assumes the form 
\be
%\begin{eqnarray}
V(v_L, v_R) = -\mu^2(v_L^2 + v_R^2)
+ (\rho_1+\rho_2)(v_L^4+v_R^4) + \rho_3v_L^2v_R^2 ~,
%\end{eqnarray}
\ee
where the parameters are inherited from the original form of
the potential \cite{mohapbook}.
Upon parametrizing $v_R=v\cos\alpha$ and $v_L=v\sin\alpha$,
the points $(v, \alpha)=(v_0, 0)$ and 
$(v_0, \pi/2)$ with $v_0=\sqrt{\mu^2/2(\rho_1+\rho_2)}$
are the minima, and 
$(\sqrt{2\mu^2/(\rho_3+2(\rho_1+\rho_2))}, \pi/4)$
a saddle point, provided $\rho_3>2(\rho_1+\rho_2)>0$. 
Electroweak phenomenology dictates that the latter condition
be valid.

It is reasonable to assume that the
effective potential continues to enjoy the above discrete
symmetry, since the same loop corrections enter for both the
fields. This means the symmetry is broken spontaneously
at the $L-R$ breaking scale, providing requisite topological condition
for the existence of domain walls. 
As the universe cools from the
$L-R$ symmetric phase, there should be causally
disconnected regions that select either $\alpha=0$ or
$\alpha=\pi/2$. Thus the vev's are functions of position 
and the two kinds of regions are separated by domain walls. 

The equations governing the wall configuration can be obtained
from minimization of the energy.
The existence of the required solution can
be shown by an extension to two variables of the standard arguments for 
one-variable solitonic solutions\cite{coleman}. 
There exists a first integral of the motion, viz., 
$h \equiv (1/2)({v'}^2+v^2{\alpha'}^2)+(-V)$ and an analysis
of equivalent problem of a point particle in an inverted
potential can be carried out. Imposing
the requirements of finiteness of total energy ensures 
that a solution exists. In particular, for 
$\rho_3 = 6 (\rho_1 + \rho_2)$, one finds an exact solution
$v_L(x)=$$(v_0/\sqrt 2)[1-\tanh(\mu x)]$;
 $v_R(x) =$$(v_0/\sqrt 2)[1+\tanh(\mu x)]$.  
This esablishes the solution at least in a neighbourhood of this
set of values of the parameters.

At the electroweak scale, the effective potential 
does not respect L-R symmetry 
due to the nature of the $\phi$ self coupling.
One finds that $v_Lv_R \sim \kappa^2$ where the symbols have been
introduced in sec. III. Upon choosing $\kappa\sim v_{EW}$ with $v_{EW}$ 
denoting the electroweak scale, $v_L$ is driven to be tiny.
The $Z_2$ guaranteeing the topological stability of the walls
now disappears. Energy minimization requires that the walls 
disintegrate. 

There is a possibility that
the L-R symmetry is not exact due to effects of a higher
unification scale,  in which case, the R  breaking 
minimum should be energetically preferred by small amounts before 
the electroweak phase transition. This will cause the
domain walls  to move around till the regions
with L breaking minimum have been converted to the true vacuum.
Some fractin of the walls would then disappear before the
electroweak scale is reached. The fate of the surviving walls
is the same as that discussed in the previous paragraph.
Further consequences are discussed in the next section.

\section{Consequences for cosmology}
From the point of view of a predictable baryogenesis,
L-R model enjoys the advantage that the primordial value of the
$B-L$ number is naturally zero, being the value of an
abelian gauge charge. 
The topological defects studied here can play a significant 
role in baryogenesis through leptogenesis.
It has been shown in \cite{mohzha} and \cite{freetal}, using 
mechanisms for electroweak baryogenesis that do not
rely on topological defects, that the parameters in the potential
require unnatural fine tuning to provide sufficient $CP$ 
violation to explain the observed asymmetry within the context of
the minimal model considered here. The cure suggested is a
singlet psuedoscalar $\sigma$ of mass either $\gg v_R$\cite{freetal} 
or $\sim v_EW$\cite{mohzha}. 

Defect mediated leptogenesis mechanisms also need this enhanced
$CP$ violation, with one exception to be noted below. For the 
present purpose we note that the $\sigma$ 
field does not alter the topological considerations presented above 
since it is a gauge singlet and
its main function is to bias the potential of the $\phi$ which
does not enter the topological defects in a significant way.
The coupling of $\sigma$ to both $\Delta_R$ and $\Delta_L$ 
may be assumed to be identical due to Left-Right symmetry.

Then the domain walls present very interesting prospects.
Their interction with other particles in the pre-electroweak 
scale plasma can result in leptogenesis.
A model independent possibility of this kind was 
considered in \cite{rbetal2}. More specific considerations also
appear in \cite{mencoo} and \cite{lewrio}. 
It is likely that the model is descended from a grand unified theory.
For this or for some other reason there may be a small
asymmetry between the L-preferring and R-preferring minima even 
above the electroweak scale. If the energy density difference 
is suppressed by powers of the GUT mass, the walls 
are still expected to be present long enough to bring about
requisite leptogenesis. 

The case of exact Left-Right symmetry leads to domain walls that
are stable before the electroweak symmetry
breaking. In this case the regions trapped in ${\vev \Delta_L} \sim v_R$
vacuum will become suddenly destabilised as the $\phi$ acquires
a vev. The destabilization can generate large amounts of entropy
and the domains should reheat to some temperature $T_H$ greater
than $v_{EW}$ but less than $v_R$. The possibility for baryogenesis
from situations with large departure from thermal equilibrium
was considered by Weinberg\cite{weinberg}. It was argued
that in such situations the asymmetry generated should be
determined by
the ratio of time constants governing baryon number violation
and entropy generation respectively. In the present case we
expect leptogenesis from the degeneration of $\vev{\Delta_L}$ 
due to the Majorana-like Yukawa coupling mentioned in sec. III.
Since the mechanism operates far from equilibrium, $CP$ violation
parameter may not enter in a significant way. The generated lepton 
asymmetry can then convert to baryon asymmetry through the 
electroweak anomaly. This possibility will be studied separately.

The cosmic strings demonstrated above can play several 
nontrivial roles in the early universe. They can provide sites
for electroweak baryogenesis as proposed in \cite{rbetal2}.
It has also been proposed that he fermion zero modes they 
possess can result in leptogenesis
\cite{rj1}.  Equally interesting is the 
process of disintegration of the unstable strings below
the electroweak scale. The decay should proceed by appearance
of gaps in the string length with formation of monopoles
at the ends of the resulting segments. The free segments then
shrink, realising the scenario of \cite{rbetal1}.
Monopoles may also catalyse lepton number creation but
their contribution should be much smaller than that of
strings or domain walls.

The Left-Right symmetric model considered here provides a
concrete setting for all of the above scenarios.
Several new features that have been demonstrated can alter
the scenarios qualitatively and merit further study.

\section*{Acknowledgement}
This work was started at the 5th workshop on High Energy 
Physics Phenomenology (WHEPP-5) which was held in IUCAA, Pune,
India
and supported by S. N. Bose National Centre for Basic Science,
Calcutta, India and Tata Institute of Fundamental Research, Mumbai,
India.  One of the authors (H.W.) thanks the University Grants 
Commission, New Delhi, for a fellowship.

\end{document}